\documentclass[aps,prb,showpacs,twocolumn,floats]{revtex4}
\usepackage{graphicx}
\usepackage{subfigure}
\usepackage{epsfig}
\usepackage{dcolumn}
\usepackage{bm}
\usepackage[ansinew]{inputenc}
\usepackage{amsmath}
\usepackage{amsthm}
\usepackage[T1]{fontenc}
\usepackage{amssymb}
\usepackage{amsfonts}
\usepackage[english]{babel}
\usepackage{enumitem}

\newcommand{\ud}{\,\mathrm{d}}
\newcommand{\ds}{\displaystyle}

\begin{document}

\title{Josephson junction with a magnetic vortex}
\date{\today}
\author{R. Zarzuela$^{1}$, E. M. Chudnovsky$^{2}$, J. Tejada$^{1}$}
\affiliation{$^{1}$Departament de F\'{i}sica Fonamental, Facultat
de F\'{i}sica, Universitat de Barcelona, Mart\'{i} i Franqu\`{e}s 1, 
08028 Barcelona, Spain\\ $^{2}$Physics Department, Lehman College,
The City University of New York, 250 Bedford Park Boulevard West,
Bronx, NY 10468-1589, U.S.A.}

\begin{abstract}
We have studied Josephson tunneling through a circularly polarized micron or submicron-size disk of a soft ferromagnetic material. 
Such a disk contains a vortex that exhibits rich classical dynamics and has recently been proposed as a tool to study quantum dynamics 
of the nanoscale vortex core. The change in the Josephson current that is related to a tiny displacement of the vortex core has been 
computed analytically and plotted numerically for permalloy disks used in experiments. It is shown that a Josephson junction with a 
magnetic disk in the vortex state can be an interesting physical system that may be used to measure the nanoscale motion of the 
magnetic vortex.
\end{abstract}


\maketitle

\section{Introduction}

Josephson tunneling through uniformly magnetized ferromagnetic layers has been intensively studied in the past \cite{Buzdin-RMP2005}. In 
this paper we are interested in the Josephson effect in the case when a ferromagnetic layer contains a vortex of the magnetization field. 
Our interest to this problem is two-fold. Firstly, micron-size disks of soft ferromagnetic materials naturally form a vortex ground state 
due to magnetic dipolar interactions.  The variety of spatial dimensions of such disks  \cite{Cowburn,Shinjo,Novosad1,Hertel} ideally suites 
typical parameters of Josephson junctions, with the disks' thickness range being comparable to the values of the coherence-length of 
conventional superconductors. 

Secondly, recent macroscopic evidence of quantum diffusion of vortices in the array of submicron and micron-size magnetic disks raises question 
whether an individual vortex tunneling event can be observed by measuring the change in the tunneling current through the disk. Curling of the 
magnetization in the plane of the disk leaves virtually no magnetic ``charges'' \cite{Feldtkeller,Usov}. Still the topology of the vortex state 
generates a very weak uncompensated magnetic moment that sticks out of the plane of the disk. It is confined to the vortex core (VC) of diameter 
comparable to the material exchange length \cite{Usov,Guslienko1}. Recent experimental works reported evidence that the dynamics of the VC is 
affected by the presence of structural defects in the sample \cite{Shima,Compton,Zarzuela3,Burgess}. This is indicative of the behavior similar 
to that of the elastic string in a random pinning potential \cite{Skvortsov}, with the finite elasticity of the vortex provided by the exchange 
interaction\cite{Zarzuela2}. 

In Ref. \onlinecite{Zarzuela3} the non-thermal magnetic relaxation under the action of an in-plane magnetic field below $T=9$ K has been 
reported. It has been attributed to the quantum diffusion of the VC in a random potential landscape towards the energy minimum. At low 
temperatures only the softest dynamical mode of the vortex can be activated, which corresponds to the gyrotropic motion of the VC. It consists 
of the circular motion of the VC \cite{Choe,Guslienko2,Guslienko3,Guslienko4,Lee} that is equivalent to the uniform precession of the magnetic 
moment of the disk. The diffusion of the VC, while conceptually similar to the quantum diffusion of an elastic string \cite{Skvortsov}, is 
mathematically different as it involves gyroscopic motion of the massless vortex \cite{Zarzuela2}. 

The problem of quantum tunneling of the VC out of the potential well created by the pinning potential has been recently studied in 
Ref. \onlinecite{Zarzuela4}. It was found, that the low-temperature quantum diffusion of the VC occurs via steps of a few interatomic distances. 
Thermal diffusion at elevated temperatures may involve longer steps. Theoretical picture of macroscopic relaxational dynamics of vortices in 
the array of micron-size permalloy disks agreed with experiment. In this paper we are asking the question whether the displacement of the VC by 
a few nanometers, or by a fraction of a nanometer, can be detected via  measurement of the tunneling current through a Josephson junction that 
is made of a magnetic disk in the vortex state. Mathematical formulation of the problem is outlined in Section \ref{sec-formulation}. The phase 
of the superconducting order parameter is calculated in Section \ref{sec-phase}. Josephson current is computed and plotted for a Py disk in 
Section \ref{sec-current}. Our results and suggestions for experiment are discussed in  Section \ref{sec-discussion}. 

\begin{figure}[htbp!]
\includegraphics[width=8cm,angle=0]{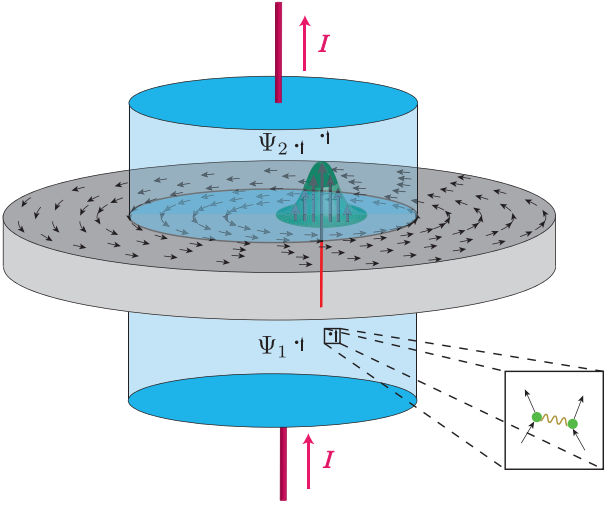}
\caption{Josephson current through a circularly polarized magnetic disk.}
\label{fig-geometry}
\end{figure}

\section{Formulation of the problem}\label{sec-formulation}

We consider a ferromagnetic Josephson junction (S/F/S), where the F-layer consists of a circularly polarized magnetic disk. This essentially 
non-uniform ground state is characterized by the curling of the magnetization in the plane of the disk and by the existence of the vortex that 
sticks out of the disk and carries small uncompensated magnetic moment,  see Fig. \ref{fig-geometry}.

Notice that, in general, ferromagnetism weakens the superconductivity at the S/F boundary due to the proximity effect. It disappears if the 
ferromagnetic and superconducting surfaces are separated by a thin non-magnetic insulating layer, leaving only electromagnetic interaction of 
the Josephson junction with the ferromagnet, which is the case studied here. Under the practical condition that the lateral size of the 
junction is smaller than the radius of the disk, but much greater than the diameter of the nanoscale VC, the Josephson current through the 
junction can be calculated rigorously. It is dominated by the configuration of the magnetization in the disk that depends on the position of 
the VC. The latter can be displaced by the external magnetic field parallel to the disk. The VC can also exhibit circular motion that 
corresponds to a collective gyroscopic mode of the disk. It can also move spontaneously via thermal or quantum diffusion in the presence of 
weak pinning. The aim of this paper is to find out whether the tiny movements of the vortex core can be detected by measuring the Josephson 
current.

The current-phase relation governing the dynamics of the Josephson effect is given by the formula \cite{Tinkham}
\begin{equation}
\label{c-ph_rel}
 j=j_{m}\sin\Phi_{21},
\end{equation}
where $\ds j_{m}=\frac{|e^{\star}|\hbar}{m^{\star}\lambda_J}|\Psi|^2$ is the maximum current density carried by the junction and 
$\Phi_{21}=\Phi_{1}-\Phi_{2}$ is the phase difference between the two superconducting regions. We assume that both superconductors 
are prepared of the same  material. Notice that $e^{\star}=-2e$ and $m^{\star}=2m_{e}$ are the charge and the mass of the Cooper pair. 
The parameter $\lambda_J$ is the property of the junction and $|\Psi|$ is the equilibrium bulk value of the modulus of the superconducting 
wave function, $\Psi = |\Psi|e^{i\Phi}$. 

In the presence of a magnetic field, the gauge-invariant phase relation is given by the formula
\begin{equation}
\label{g-i_ph_rel}
\Phi_{21}=\Phi^{(0)}_{21}+\frac{2\pi}{\Phi_{0}}\int_{1}^{2}\vec{A}\cdot\ud\vec{l},
\end{equation}
where $\Phi^{(0)}_{21}$ is the phase difference across the junction, $\ds \Phi_{0}={h c}/{|e^{\star}|}$ is the flux quantum, and $\vec{A}$ is 
the vector potential that is determined by the magnetization field $\vec{M}$ within the volume $V$ of the disk, 
\begin{equation}
 \vec{A}(\vec{r})=\int_{V}\frac{\nabla'\times\vec{M}(\vec{r}\,')}{|\vec{r}-\vec{r}\,'|}\ud^{3}\vec{r}\,'+\oint_{\partial V}\frac{\vec{M}(\vec{r}\,')
 \times\vec{n}'}{|\vec{r}-\vec{r}\,'|}\ud^{2}S'.
\end{equation}
Here $\vec{n}'$ is the vector normal to the surface of the disk. 

Let $L$ and $R$ be respectively the thickness and the radius of the ferromagnetic disk. We set the coordinate frame according to the symmetry of 
the system: the $XY$ plane coincides with the plane of the disk and the $Z$ axis coincides with the symmetry axis of the disk. The S/F boundaries 
are located at $z=\pm L/2$.
According to the geometry of our system, the path integral in Eq. \eqref{g-i_ph_rel} must be performed along the $Z$ axis between $z=-L/2$ 
(superconducting region 1/ferromagnet boundary) and $z=L/2$ (ferromagnet/superconducting region 2 boundary). Therefore, the gauge-invariant 
phase relation becomes 
\begin{equation}
\label{g-i_ph_rel_bis}
\Phi_{21}=\Phi^{(0)}_{21}+\frac{2\pi}{\Phi_{0}}\int_{-\frac{L}{2}}^{\frac{L}{2}}A_{z}(\vec{r})\ud z,
\end{equation}
where $A_{z}$ is the projection of the vector potential onto the $Z$-axis. 

\section{Computation of the phase difference}\label{sec-phase}

\begin{figure}[htbp!]
\includegraphics[width=10cm,angle=0]{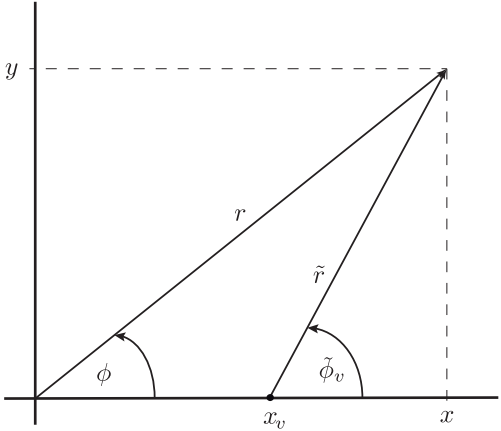}
\caption{Relation between two systems of polar coordinates used in the text.}
\label{fig-polar}
\end{figure}

The magnetization field in the disk can be described by the fixed-length vector
\begin{eqnarray}
&& \vec{M}(\Theta,\Phi)=M_{s}(\cos\Phi\sin\Theta,\sin\Phi\sin\Theta,\cos\Theta) \\ \nonumber
&& =M_{s}(\sqrt{1-m^2}\cos\Phi,\sqrt{1-m^2}\sin\Phi,m),
\end{eqnarray}
where $M_{s}$ is the saturation magnetization of the ferromagnetic material and $m=\cos\Theta$ is the projection of the normalized magnetic 
moment onto the $z$ axis. Let $\vec{X}_{v}(t,z)=\Big(x_{v}(t,z),y_{v}(t,z)\Big)$ be the coordinates of the center of the VC in the $XY$ plane. 
We assume the rigidity of the vortex structure, which translates into the VC coordinates being independent of the $Z$ variable. We use a 
quasi-static approximation in which no time dependence of the VC coordinates is considered, which is always valid for the slow motion of the 
vortex. Because of this, we can rotate the coordinate axis in the $XY$ plane so that $\vec{X}_{v}=x_{v}\hat{e}_{x}$. Let $(r,\phi)$ be the polar 
coordinates in the $XY$ plane. The static solution of the magnetization field is\cite{Zarzuela2}
\begin{align}
\label{zeroth}
\Phi_{0}(x,y)&=\tan^{-1}(y/x-x_{v})+\phi_{0}\nonumber\\
\cos\Theta_{0}(\tilde{r})&= \left\{\begin{array}{lcc}
            p\left(1-C_{1}\left(\frac{\tilde{r}}{\Delta_{0}}\right)^2\right)
            & & \tilde{r} \ll\Delta_{0}\\
        C_{2}\left(\frac{\Delta_{0}}{\tilde{r}}\right)^{1/2}\exp(-\tilde{r}/\Delta_{0})
        & & \tilde{r}\gg\Delta_{0}
           \end{array}\nonumber\right.
\end{align}
where 
\begin{itemize}
\item[$\bullet$] $\tilde{r}=||\vec{r}-\vec{X}_{v}||_{2}$ is the radial distance from the VC center.
\item[$\bullet$] $p=\pm1$ is the polarization of the magnetization. 
\item[$\bullet$] $\phi_{0}=\pm\pi/2$ corresponds to counter-clockwise/ clockwise rotation of the magnetization. 
\item[$\bullet$] $\Delta_{0}=\sqrt{A/M_{s}^2}$ is the exchange length of the material.
\item[$\bullet$] $C_{1}=\frac{3}{7}$ and $C_{2}=\frac{4}{7}pe$.
\end{itemize}
Introducing the vector basis for cylindrical coordinates,
\begin{equation}
 \hat{e}_{r}=\cos\phi \hat{e}_{x}+\sin\phi\hat{e}_{y},\quad \hat{e}_{\phi}=-\sin\phi\hat{e}_{x}+\cos\phi\hat{e}_{y},
\end{equation}
the magnetization field can be split into $\vec{M}(r,\phi,z)=M_{r}\hat{e}_{r}+M_{\phi}\hat{e}_{\phi}+M_{z}\hat{e}_{z}$, with
\begin{eqnarray}
 M_{r}&=&\hat{e}_{r}\cdot\vec{M}=M_{s}\sqrt{1-m_{0}^2}\cos(\Phi_{0}-\phi)\\
 M_{\phi}&=&\hat{e}_{\phi}\cdot\vec{M}=M_{s}\sqrt{1-m_{0}^2}\sin(\Phi_{0}-\phi)\\
 M_{z}&=& M_{s} m_{0}
\end{eqnarray}
Let $\tilde{\phi}_{v}=\tan^{-1}(y/x-x_{v})$. Then $(\tilde{r},\tilde{\phi}_{v})$ are the polar coordinates in the $XY$ plane from the VC center. 
Fig. \ref{fig-polar} shows the geometrical relation between both systems of polar coordinates, from which we straightforwardly deduce the 
following identities,
\begin{align}
\label{eqs1}
&\tilde{r}=||\vec{r}-\vec{X}(t,z)||_{2}=\sqrt{r^2+x_{v}^{2}-2r x_{v}\cos\phi},\\
&\sin\tilde{\phi}_{v}=\frac{r}{\tilde{r}}\sin\phi,\quad\cos\tilde{\phi}_{v}=\frac{1}{\tilde{r}}(r\cos\phi-x_{v})\nonumber
\end{align}

According to the asymptotic dependences of the static solution we have
\begin{align}
\label{eqs2}
\sqrt{1-m_{0}^2}&\simeq \left\{\begin{array}{lcc}
            \sqrt{2C_{1}}\frac{\tilde{r}}{\Delta_{0}} & & \tilde{r} \ll\Delta_{0}\\
            1 & & \tilde{r}\gg\Delta_{0}
           \end{array}\right.
\end{align}
and
\begin{equation}
\label{eqs3}
\sin\Phi_{0}=C\cos\tilde{\phi}_{v},\qquad \cos\Phi_{0}=-C\sin\tilde{\phi}_{v},
\end{equation}
where $C=\sin\phi_{0}=\pm1$ represents the circulation of magnetization field of the ground state.

In this paper we consider the limit $R\gg L,\Delta_{0}$. Being interested in the tiny displacements of the VC due to, e.g., quantum tunneling, 
we also shall assume that $|x_{v}|\ll\Delta_{0}$. This allows us to obtain a perturbative expansion of the phase difference across the junction 
in terms of powers of $x_{v}$. 

\subsection{Surface contribution}

The surface of the disk consists of three surfaces, $\partial V=S_1\cup S_2\cup S_3$, where $S_1$ and $S_3$ are respectively the top and the 
bottom surfaces of the disk, and $S_2$ is the lateral surface. The corresponding normal vectors are $\hat{n}_{1}=-\hat{n}_{3}=\hat{e}_{z}$ and 
$\hat{n}_{2}=\hat{e}_{r}$. It is straightforward to prove the following identities
\begin{eqnarray}
 &&\vec{M}\times\hat{n}|_{S_1}=-\vec{M}\times\hat{n}|_{S_3}=M_{\phi}\hat{e}_{r}-M_{r}\hat{e}_{\phi} \\
&& \vec{M}\times\hat{n}|_{S_2}= M_{z}\hat{e}_{\phi}-M_{\phi}\hat{e}_{z},
\end{eqnarray}
so that the surface contribution to $A_{z}$ comes from integration over $S_{2}$. This means that 
\begin{eqnarray}
&& A_{z}(\vec{r})|_{\textrm{Surf}}=\int_{S_2}\hat{e}_{z}\cdot \frac{\vec{M}(\vec{r}\,')
 \times\vec{n}'}{|\vec{r}-\vec{r}\,'|}\ud^{2}S' = \\ \nonumber
&&\int_{S_2}r'\ud\phi'\ud z' \frac{-M_{\phi}}{|\vec{r}-\vec{r}\,'|}=
 R\int_{-\frac{L}{2}}^{\frac{L}{2}}\ud z'\int_{0}^{2\pi}\ud\phi'\frac{-M_{\phi}}{|\vec{r}-\vec{r}\,'|}\bigg|_{S_{2}}
\end{eqnarray}
The Coulomb potential can be expanded in cylindrical coordinates as
\begin{equation}
 \label{Coulomb}
 \frac{1}{|\vec{r}-\vec{r}\,'|}=\int_{0}^{\infty}\ud k\; J_{0}(k\tilde{r})e^{-k(z_{>}-z_{<})},
\end{equation}
were $z_{<}=\min\{z,z'\}$, $z_{>}=\max\{z,z'\}$, $J_{0}(x)$ is the zero-order Bessel function of the first kind, and
\begin{equation}
\tilde{r} = \sqrt{r^2+r'^2-2rr'\cos(\phi-\phi')}
\end{equation}
If $z\neq z'$ we can switch to the
integration over $\phi$ and $k$, which gives
\begin{align}
\label{switch}
\int_{0}^{2\pi}&\ud\phi'(-M_{\phi}|_{S_{2}})\int_{0}^{\infty}\ud k\; J_{0}(k\tilde{r})e^{-k(z_{>}-z_{<})}=\\
&\int_{0}^{\infty}\ud k\;e^{-k(z_{>}-z_{<})}\int_{0}^{2\pi}\ud\phi' J_{0}(k\tilde{r})(-M_{\phi}|_{S_{2}}),\nonumber
\end{align}

By means of Eqs. \eqref{eqs1},\eqref{eqs2} and \eqref{eqs3} we obtain the following asymptotic expressions for $M_{\phi}$,
\begin{align}
M_{\phi}&\simeq \left\{\begin{array}{lcc}
            M_{s}\frac{C}{\Delta_{0}}\sqrt{2C_{1}}\left[r'-x_{v}\cos\phi'\right] & & \tilde{r}' \ll\Delta_{0}\\
            \frac{M_{s} C}{\tilde{r}'}\left[r'-x_{v}\cos\phi'\right] & & \tilde{r}'\gg\Delta_{0}
           \end{array}\right.
\end{align}
Integration over surface $S_{2}$ corresponds to the asymptotic limit $\tilde{r}'\gg\Delta_{0}$, which leads to $r'\gg x_{v}$. Consequently, we 
can use the following expansion of the Coulomb potential
\begin{equation}
 \label{radial}
 \frac{1}{\tilde{r}'}\simeq\frac{1}{r'}+\frac{x_{v}\cos\phi'}{r'^2}
\end{equation}
and $M_{\phi}|_{S_{2}}\simeq M_{s}C\left((1+O(x_{v}/r))^2\right)$.

Neumann's addition theorem for Bessel functions leads to the following expansion
\begin{equation}
\label{Neumann}
J_{0}(k\tilde{r}) =\sum_{m\geq0}\epsilon_{m}J_{m}(kr)J_{m}(kr')\cos\left(m(\phi-\phi')\right),
\end{equation}
where $\epsilon_{0}=1$ and $\epsilon_{m}=2,\;m>0$. With account of the orthogonality of the Fourier basis 
$\{1\}\cup\{\cos m\phi'\}_{m\in\mathbb{N}}\cup\{\sin m\phi'\}_{m\in\mathbb{N}}$ one obtains
\begin{align}
\label{eqs4}
\int_{0}^{2\pi}\ud\phi' &J_{0}(k\tilde{r})(-M_{\phi})|_{S_{2}}=\\
&-2\pi M_{s}C J_{0}(kr)J_{0}(kR)+O\left((x_{v}/r)^2\right).\nonumber
\end{align}
On the other hand, we have the identity 
\begin{eqnarray}
\label{Legendre}
 &&\int_{0}^{\infty}\ud k J_{m}(kr)J_{m}(kr')e^{-k(z_{>}-z_{<})} \\ \nonumber
&&=\frac{1}{\pi\sqrt{rr'}}Q_{m-\frac{1}{2}}\left[\frac{r^2+r'^2+(z-z')^2}{2rr'}\right],
\end{eqnarray}
where $Q_{\lambda}[z]$ is the Legendre function of second kind of the degree $\lambda$,
\begin{eqnarray}
&& Q_{\lambda}[z]=\frac{\sqrt{\pi}}{2^{\lambda+1}}\frac{\Gamma(\lambda+1)}{\Gamma(\lambda+3/2)}\frac{1}{z^{\lambda+1}} \times \\ \nonumber
&& {}_2F_1\left(
 \frac{\lambda+1}{2},\frac{\lambda}{2}+1,\lambda+\frac{3}{2};\frac{1}{z^2}\right),
\end{eqnarray}
with $_2 F_1$ being the hypergeometric function. Consequently, Eqs. \eqref{Coulomb},\eqref{eqs4} and \eqref{Legendre} give
\begin{align}
&\int_{0}^{2\pi}\ud\phi'\frac{-M_{\phi}}{|\vec{r}-\vec{r}	\,'|}\bigg|_{S_{2}}=-\frac{2 M_{s}C}{\sqrt{rR}}\times\\
&Q_{-\frac{1}{2}}\left[\frac{r^2+R^2+(z-z')^2}{2rR}\right]+O\left((x_{v}/r)^2\right),\quad z\neq z'.\nonumber
\end{align}

The contribution of the ferromagnetic layer to the phase difference of the junction is given by the path integral [see 
Eq. \eqref{g-i_ph_rel_bis}]
\begin{equation}
 \Phi_{21}^{F}|_{\textrm{Surf}}= \frac{2\pi R}{\Phi_{0}}\int_{-\frac{L}{2}}^{\frac{L}{2}}\ud z\int_{-\frac{L}{2}}^{\frac{L}{2}}\ud z'\int_{0}^{2\pi}\ud\phi'\frac{-M_{\phi}}{|\vec{r}-\vec{r}\,'|}\bigg|_{S_{2}},
\end{equation}
To deal with the singularity of the integrand when $\vec{r}\,'$ equals $\vec{r}$, we introduce the Cauchy principal value prescription to the 
integration over the $z$ variable, that is
\begin{equation}
 \int_{-\frac{L}{2}}^{\frac{L}{2}}\ud z'\Rightarrow \mathcal{P}\int_{-\frac{L}{2}}^{\frac{L}{2}}\ud z':=
 \lim_{\epsilon\rightarrow0^{+}}\left\{\int_{-\frac{L}{2}}^{z-\epsilon}\ud z'+\int_{z+\epsilon}^{\frac{L}{2}}\ud z'\right\}.
\end{equation}
With account of this prescription Eq. \eqref{switch} can be always applied and so $\Phi_{21}^{F}|_{\textrm{Surf}}$ becomes
\begin{eqnarray}
\label{surf_ph_diff}
 &&\Phi_{21}^{F}|_{\textrm{Surf}}=-2 M_{s}C\frac{2\pi}{\Phi_{0}}\sqrt{\frac{R}{r}}\int_{-\frac{L}{2}}^{\frac{L}{2}}\ud z \times \;\\
 &&\mathcal{P}\int_{-\frac{L}{2}}^{\frac{L}{2}}\ud z'Q_{-\frac{1}{2}}\left[\frac{r^2+R^2+(z-z')^2}{2rR}\right]+O(x_{v}^2)\nonumber
\end{eqnarray}

\subsection{Bulk contribution}

Bulk contribution to the phase difference of the Josephson junction stems from the projection of the curl of the magnetization field onto the 
$Z$ axis. That is,
\begin{eqnarray}
&& A_{z}(\vec{r})|_{\textrm{Bulk}}=\int_{V}\frac{\hat{e}_{z}\cdot\left(\nabla'\times\vec{M}(\vec{r}\,')\right)}{|\vec{r}-\vec{r}\,'|}\ud^{3}\vec{r}\,'\\ \nonumber
&&=\int_{-\frac{L}{2}}^{\frac{L}{2}}\ud z'\int_{0}^{R}\ud r'\int_{0}^{2\pi}\ud\phi'\frac{r'\left(\nabla'\times\vec{M}(\vec{r}\,')\right)}{|\vec{r}-\vec{r}\,'|}\cdot\hat{e}_{z}
\end{eqnarray}
with the projection of $\nabla'\times\vec{M}(\vec{r}\,')$ onto the $Z$ axis being 
\begin{equation}
 \hat{e}_{z}\cdot\left(\nabla'\times\vec{M}(\vec{r}\,')\right)=\frac{1}{r'}\left(\frac{\partial (r' M_{\phi'})}{\partial r'}-\frac{\partial M_{r'}}{\partial\phi'}\right).
\end{equation}
As in the previous section, with account of Eq. \eqref{Coulomb} for the cylindrical expansion of the Coulomb potential (if $z\neq z'$) we have 
\begin{align}
\label{switch_bulk}
&\int_{0}^{2\pi}\ud\phi'\left[r' \hat{e}_{z}\cdot\left(\nabla'\times\vec{M}\right)\right]\int_{0}^{\infty}\ud k\; J_{0}(k\tilde{r})e^{-k(z_{>}-z_{<})}= \nonumber\\
&\int_{0}^{\infty}\ud k\;e^{-k(z_{>}-z_{<})}\int_{0}^{2\pi}\ud\phi' J_{0}(k\tilde{r})\left[r' \hat{e}_{z}\cdot\left(\nabla'\times\vec{M}\right)\right],
\end{align}
Let $\bar{z}^{(}{}'{}^{)}=z^{(}{}'{}^{)}/\Delta_{0}$ and $\rho^{(}{}'{}^{)}=r^{(}{}'{}^{)}/\Delta_{0}$ be the set of normalized spatial 
coordinates. With account of the normalized versions of Eqs. \eqref{eqs1},\eqref{eqs2},\eqref{eqs3},\eqref{radial}, and of the approximation 
$\tilde{\rho}'\simeq\rho'$ in the asymptotic regime $\tilde{\rho}'\ll 1$,  we have the following asymptotic expressions
\begin{align}
 \hat{e}_{z}\cdot\Big(\nabla'\times\vec{M}\Big)\simeq&\frac{M_{s}C\sqrt{2C_{1}}}{\Delta_{0}}\Big[(2-C_{1}\rho'^2)-2\frac{x_{v}}{\Delta_{0}\rho'}\\
 &\times(2-C_{1}\rho'^2)\cos\phi'\Big]\nonumber
\end{align}
for $\tilde{\rho}'\ll1$ and 
\begin{align}
 \hat{e}_{z}\cdot\left(\nabla'\times\vec{M}\right)\simeq\frac{M_{s} C}{\Delta_{0}\rho'}\left[1+\frac{x_{v}}{\Delta_{0}\rho'}\cos\phi'\right]
\end{align}
in the asymptotic regime $\tilde{\rho}'\gg1$.

Again, in the case of $z\neq z'$ the addition theorem \eqref{Neumann}, orthogonality of the Fourier basis and the identity \eqref{Legendre} lead to the following 
asymptotic expressions 
\begin{align}
\label{Bulk_asym1}
&\int_{0}^{2\pi}\ud\phi'\frac{r'\left(\nabla'\times\vec{M}\right)\cdot\hat{e}_{z}}{|\vec{r}-\vec{r}\,'|}\simeq\frac{2M_{s}C\sqrt{2C_{1}}}{\Delta_{0}\sqrt{\rho\rho'}}\times\\
&\left[(2-C_{1}\rho'^2)\rho'Q_{-\frac{1}{2}}[\chi]-2\frac{x_{v}}{\Delta_{0}}(2-C_{1}\rho'^2)Q_{\frac{1}{2}}[\chi]\cos\phi\right]\nonumber
\end{align}
for $\tilde{\rho}'\ll1$ and 
\begin{align}
\label{Bulk_asym2}
\int_{0}^{2\pi}\ud\phi'\frac{r'\left(\nabla'\times\vec{M}\right)}{|\vec{r}-\vec{r}\,'|}&\cdot\hat{e}_{z}\simeq\frac{2M_{s} C}{\Delta_{0}\sqrt{\rho\rho'}}\times\\
&\Big[Q_{-\frac{1}{2}}[\chi]+\frac{x_{v}}{\Delta_{0}\rho'}Q_{\frac{1}{2}}[\chi]\cos\phi\Big]\nonumber
\end{align}
in the asymptotic regime $\tilde{\rho}'\gg1$, where the expansions in the right side have been performed up to first order in the VC 
displacement and
\begin{equation}
\chi=\frac{\rho^2+\rho'^2+(\bar{z}-\bar{z}')^2}{2\rho\rho'}.
\end{equation}

In the same manner as in the previous section, we introduce the Cauchy principal value prescription to avoid singularities in the integrand of 
the bulk contribution to the phase difference. Therefore we obtain 
\begin{align}
\label{Bk_ph_diff1}
 &\Phi_{21}^{F}|_{\textrm{Bulk}}=\frac{2\pi}{\Phi_{0}}\int_{\frac{-L}{2}}^{\frac{L}{2}}A_{z}(\vec{r})|_{\textrm{Bulk}}\ud z=\frac{2\pi}{\Phi_{0}}\Delta_{0}^{3}\int_{\frac{-L}{2\Delta_{0}}}^{\frac{L}{2\Delta_{0}}}\mathrm{d} \bar{z}\nonumber\\
 &\times\mathcal{P}\int_{\frac{-L}{2\Delta_{0}}}^{\frac{L}{2\Delta_{0}}}\ud \bar{z}'\int_{0}^{\frac{R}{\Delta_{0}}}\ud \rho' \int_{0}^{2\pi}\ud\phi'\frac{r'\left(\nabla'\times\vec{M}\right)\cdot\hat{e}_{z}}{|\vec{r}-\vec{r}\,'|}\nonumber\\
 &=\frac{2\pi}{\Phi_{0}}\frac{2M_{s}C\Delta_{0}^{2}}{\sqrt{\rho}}\int_{\frac{-L}{2\Delta_{0}}}^{\frac{L}{2\Delta_{0}}}\ud \bar{z}\;\mathcal{P}\int_{\frac{-L}{2\Delta_{0}}}^{\frac{L}{2\Delta_{0}}}\ud \bar{z}'\Bigg\{\sqrt{2C_{1}}
 \int_{0}^{1}\frac{\ud\rho'}{\sqrt{\rho'}}\nonumber\\
 &\times\Big[(2-C_{1}\rho'^2)\rho'Q_{-\frac{1}{2}}[\chi]-2\frac{x_{v}}{\Delta_{0}}(2-C_{1}\rho'^2)Q_{\frac{1}{2}}[\chi]\cos\phi\Big]\nonumber\\
 &+\int_{1}^{\frac{R}{\Delta_{0}}}\frac{\ud\rho'}{\sqrt{\rho'}}\; \left[Q_{-\frac{1}{2}}[\chi]+\frac{x_{v}}{\Delta_{0}\rho'}Q_{\frac{1}{2}}[\chi]\cos\phi\right]\Bigg\},
\end{align}
where we have split integration over $\rho'$ into the domains $[0,1]$ and $[1,R/\Delta_{0}]$ corresponding to the asymptotic expansions of the 
integrand [see Eq. \eqref{Bulk_asym1} and \eqref{Bulk_asym2}]. As before, we are working under the assumption of an infinitesimal displacement 
of the VC from the center of the disk, $x_{v}\ll1$, so that the deformation of the VC area with respect to the centered case ($\rho'\leq1$) is 
small and can be safely neglected in the integration process, simplifying the calculations. 

\section{Computation of the Josephson current}\label{sec-current}

According to the gauge-invariant phase relation [Eq. \eqref{g-i_ph_rel}], the superconducting phase difference splits into the sum of the 
intrinsic component and of both surface and bulk contributions, which are given by Eqs. \eqref{surf_ph_diff} and \eqref{Bk_ph_diff1} 
respectively, due to the presence of the F-layer. That is,
\begin{align}
\Phi_{21}(\rho,\phi)&=\Phi_{21}^{0}+\Phi_{21}^{F}|_{\textrm{Surf}}+\Phi_{21}^{F}|_{\textrm{Bulk}}\\
&=\Phi_{21}^{0}+a(\rho)+b(\rho)\cos\phi,\nonumber
\end{align}
where the functions $a(\rho)$ and $b(\rho)$ are given by
\begin{align}
&a(\rho)=\frac{4\pi M_{s}C\Delta_{0}^{2}}{\Phi_{0}\sqrt{\rho}}\int_{\frac{-L}{2\Delta_{0}}}^{\frac{L}{2\Delta_{0}}}\ud \bar{z}\;\mathcal{P}\int_{\frac{-L}{2\Delta_{0}}}^{\frac{L}{2\Delta_{0}}}\ud \bar{z}'\times\\
&\Bigg\{\sqrt{2C_{1}}\int_{0}^{1}\frac{\ud\rho'}{\sqrt{\rho'}}(2-C_{1}\rho'^2)\rho'Q_{-\frac{1}{2}}[\chi]+\int_{1}^{\frac{R}{\Delta_{0}}}\frac{\ud\rho'}{\sqrt{\rho'}}\; Q_{-\frac{1}{2}}[\chi]\nonumber\\
&-\sqrt{\frac{R}{\Delta_{0}}}Q_{-\frac{1}{2}}\left[\frac{\rho^2+(R/\Delta_{0})^2+(\bar{z}-\bar{z}')^2}{2\rho R/\Delta_{0}}\right]\Bigg\}.\nonumber\\
&b(\rho)=\frac{x_{v}}{\Delta_{0}}\frac{4\pi M_{s}C\Delta_{0}^{2}}{\Phi_{0}\sqrt{\rho}}\int_{\frac{-L}{2\Delta_{0}}}^{\frac{L}{2\Delta_{0}}}\ud \bar{z}\;\mathcal{P}\int_{\frac{-L}{2\Delta_{0}}}^{\frac{L}{2\Delta_{0}}}\ud \bar{z}'\times\\
&\Bigg\{\int_{1}^{\frac{R}{\Delta_{0}}}\frac{\ud\rho'}{\rho'^{3/2}}\; Q_{\frac{1}{2}}[\chi]-2\sqrt{2C_{1}}\int_{0}^{1}\frac{\ud\rho'}{\sqrt{\rho'}}(2-C_{1}\rho'^2)Q_{\frac{1}{2}}[\chi]\Bigg\}.\nonumber
\end{align}

The current density across the junction is given by the current-phase relation \eqref{c-ph_rel}:
\begin{equation}
\label{current}
j=j_{m}\sin\big(\Phi_{21}^{0}+a(\rho)+b(\rho)\cos\phi\big).
\end{equation}
With account of the Jacobi-Anger expansions 
\begin{align}
\cos(b(\rho)\cos\phi)&=J_{0}(b(\rho))+2\sum_{n\geq1}(-1)^{n} J_{2n}(b(\rho))\cos(2n\phi),\nonumber\\
\sin(b(\rho)\cos\phi)&=2\sum_{n\geq0}(-1)^{n} J_{2n+1}(b(\rho))\cos\big((2n+1)\phi\big),
\end{align}
and of the uniform convergence of these series ($b(\rho)$ is a bounded function over the domain $[0,R/\Delta_{0}]$), we have the identities
\begin{eqnarray}
\int_{0}^{2\pi}\ud\phi\;\sin\big(b(\rho)\cos\phi\big)&=&0,\\
\int_{0}^{2\pi}\ud\phi\;\cos\big(b(\rho)\cos\phi\big)&=&2\pi J_{0}(b(\rho)).
\end{eqnarray}
Combined with the trigonometric identity $\sin(\Phi_{21}^{0}+a(\rho)+b(\rho)\cos\phi)=\sin\big(\Phi_{21}^{0}+a(\rho)\big)\cos(b(\rho)\cos\phi)+\cos\big(\Phi_{21}^{0}+a(\rho)\big)\sin(b(\rho)\cos\phi)$ they lead to the following expression 
\begin{align}
\int_{0}^{2\pi}\ud\phi\;\sin\big(\Phi_{21}^{0}+a(\rho)&+b(\rho)\cos\phi\big)=\\
&2\pi J_{0}\big(b(\rho)\big)\sin\big(\Phi_{21}^{0}+a(\rho)\big).\nonumber
\end{align}

The total current can be obtained by integrating Eq. \eqref{current} over the surface, $S_J = \pi R_0^2$, of the junction of radius $R_0$, 
centered at the origin of the $XYZ$ coordinate frame.  Therefore, we obtain the following expression for the total current
\begin{align}
I&=\int_{S_{J}}\ud^{2}\vec{r}\; j_{m}\sin\Phi_{21}=\\
&=I_{m}\int_{0}^{R_{0}/\Delta_{0}}\rho\ud\rho\; J_{0}\big(b(\rho)\big)\sin\big(\Phi_{21}^{0}+a(\rho)\big),\nonumber
\end{align}
where 
\begin{equation}
\ds I_{m}=2\pi j_{m}\Delta_{0}^2=\frac{|e^{\star}|h}{m^{\star}\lambda_J}|\Psi|^2\Delta_{0}^2
\end{equation}
is the maximum current carried by the junction.  

\begin{figure}[htbp!]
\includegraphics[width=8.5cm, angle=0]{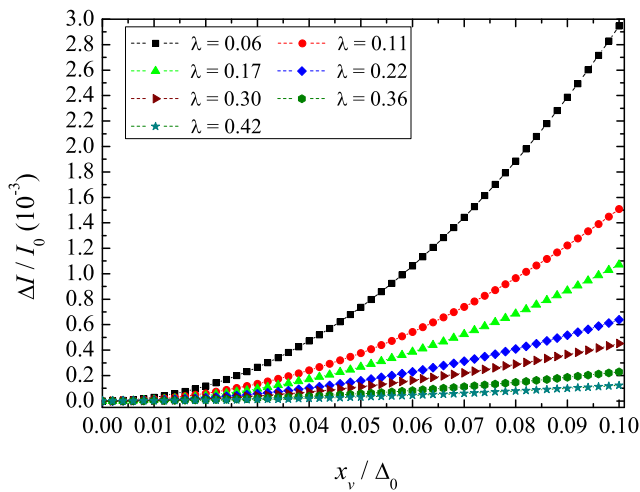}
\caption{Variation of the Josephson current for different values of the parameter $\lambda=R_0/R$ as a function of the normalized VC 
displacement from the center of the disk. The current is normalized by $I_{0}=I(x_{v}=0)$.}
\label{fig-current}
\end{figure}

In estimating the effect of the displacement of the VC we shall assume that the intrinsic phase difference of the junction, $\Phi_{21}^{0}$, 
is zero. Fig. \ref{fig-current} shows variation of the Josephson current, $\Delta I = I_{0}-I$, resulting from small displacements of the VC 
with respect to the center of the disk, for different values of the parameter $\lambda=R_{0}/R$. The variation of the current is normalized 
with respect to $I_{0}=I(x_{v}=0)$. Computation has been performed for a permalloy disk of normalized radius $R/\Delta_{0}=100$ and normalized 
thickness $L/\Delta_{0}=6$. We have used the experimental values $M_{s}=7.5\cdot10^5$ A/m and $A=1.3\cdot10^{-11}$ J/m for permalloy, which give 
$\Delta_{0}\simeq15.2$ nm.

\section{Discussion}\label{sec-discussion}

We have studied how the tunneling current through the Josephson junction containing a circularly polarized magnetic disk changes when the center 
of the vortex is displaced by a tiny distance due to, e.g, thermal activation or quantum tunneling. The numerical work has been done for disk of 
the thickness that is few times greater than the diameter of the vortex core. The latter in permalloy is about $15$nm, which for 
$L/\Delta_0 = 6$ used in the plot of Fig. \ref{fig-current} corresponds to the disks of thickness of $90$nm. Such disks have been experimented 
with in Refs. \onlinecite{Zarzuela3,Zarzuela2} where thermal and quantum diffusion of vortices has been observed. Such a thickness of the disk 
is well suited for the use in the Josephson junction.

The change in the Josephson current due to the displacement of the VC from the center of the disk has been computed in the range up to 
$x_v \sim 0.1\Delta_0$, which corresponds to $1.5$nm for a permalloy disk. The maximal change in the Josephson current in this range of the 
displacement is of order of a few tenth of a percent, which is within experimental range. It grows fast with the displacement for $R_0/R$ below 
$0.1$, which for a disk of radius of $1.5\mu$m used in  Refs. \onlinecite{Zarzuela3,Zarzuela2} corresponds to the Josephson junction of the 
lateral size $0.3\mu$m. While in calculations we used a circular junction, its geometry does not really matter as long as its size $R_0 < R$ is 
large compared to $\Delta_0$.  Smaller junctions produce stronger effect. 

Our calculation and numerical estimates clearly illustrate that a Josephson junction with a magnetic disk in the vortex state would be an 
interesting physical system that can be used to measure the nanoscale motion of the vortex core. Manufacturing of such junctions and 
experimenting with them may open up an exciting field of research on quantum and classical dynamics of magnetic vortices. 

\section{Acknowledgements}

The work at the University of Barcelona was supported by the Spanish Government Project No. MAT2008-04535. 
The work of E.M.C. at Lehman College is supported by the Department of Energy through grant No. DE-FG02-93ER45487.

\end{document}